\newcommand{\bra}[1]{\ensuremath{\left\langle {#1} \right|}}
\newcommand{\ket}[1]{\ensuremath{\left|  #1 \right\rangle}}

\documentclass[12pt]{article}
\usepackage{graphicx}
\usepackage{fixltx2e}
\usepackage{amsmath}

\usepackage{scicite}

\usepackage{times}

\newenvironment{sciabstract}{%
\begin{quote} \bf}
{\end{quote}}

\newcounter{lastnote}

\title{Reduced back-action for phase sensitivity 10 times beyond the standard quantum limit}

\author
{Justin G. Bohnet$^{1}$,  Kevin C. Cox$^{1}$, Matthew A. Norcia$^{1}$,  Joshua M. Weiner$^{1}$, \\
Zilong Chen$^{1}$, \& James K. Thompson$^{1\ast}$ \\
\\
\normalsize{$^{1}$JILA, University of Colorado at Boulder and NIST,}\\
\normalsize{Boulder, Colorado 80309, USA}\\
\\
\normalsize{$^\ast$To whom correspondence should be addressed; E-mail:  jkt@jila.colorado.edu.}
}

\date{}

\begin{document} 


\baselineskip24pt


\maketitle


\begin{sciabstract}

Collective measurements can project a system into an entangled state with enhanced sensitivity for measuring a quantum phase,  but measurement back-action has limited previous efforts to only modest improvements. Here we use a collective measurement to produce and directly observe, with no background subtraction, an entangled, spin-squeezed state with phase resolution improved in variance by a factor of 10.5(1.5), or 10.2(6) dB, compared to the initially unentangled ensemble of $N = 4.8\times10^5$ $^{87}$Rb atoms. The measurement uses a cavity-enhanced probe of an optical cycling transition to mitigate back-action associated with state-changing transitions induced by the probe. This work establishes collective measurements as a powerful technique for generating entanglement for precision measurement, with potential impacts in biological sensing, communication, navigation, and tests of fundamental physics.

\end{sciabstract}



A defining characteristic of quantum mechanics is the ability of a measurement to change the state of the system being measured. 
For example, a measurement of a system in a superposition of two states causes the system to project, or collapse, into one of the two discrete states.
Measurements performed on an ensemble, however, can project the ensemble into an entangled state when only collective quantities are measured. 
For instance, here we measure a cavity field that is entangled with the total number of spin-1/2 atoms in spin up (Fig. 1A). 
Any information about the spin-state of a single atom that leaks to the environment due to imperfections in the collective measurement reduces entanglement due to collapse of individual atoms. 
Such collective or joint measurements arise in a wide range of applications, including quantum teleportation\cite{olmschenk2009}, quantum information protocols\cite{Kok07OpticalQuantumComputing}, studies of strongly-correlated quantum systems\cite{Eckert2007}, Dicke superradiance\cite{DIC54}, and entanglement generation in optical\cite{Guerlin2007}, solid state\cite{RDW2013} and atomic systems\cite{chou2005measurement}.
  
Entanglement generated by a collective measurement can be used to overcome the fundamental quantum randomness that limits a diverse set of precision measurements\cite{kuzmich98}. 
Atomic sensors in particular are nearly or already limited by quantum noise, so entanglement-enhanced metrology would improve some of the most precise measurements of external fields\cite{Kominis2003}, rotations\cite{GBK97}, and time\cite{Hinkley2013}, and will advance searches for new physics\cite{Orzel2012}.
Atomic sensors encode their information in a quantum phase $\theta$, whose value is estimated by measuring the population of atoms in different quantum states. Quantum projection noise \cite{Itano1993} for an ensemble of $N$ independent atoms limits the uncertainty in the estimate of $\theta$ to a variance $\Delta \theta^2 \ge \Delta \theta_{SQL}^2 = 1/N$, a limit known as the standard quantum limit (SQL) for a coherent spin state (CSS). Entanglement can be used to bypass this limitation in atomic sensors, as well as in microwave\cite{Eichler2011} and optical\cite{Eberle2010} fields.

A collective measurement that both resolves the quantum noise that appears in $\theta$ and induces sufficiently small measurement back-action can be used to subtract quantum noise from subsequent measurements of $\theta$\cite{kuzmich98}. While the resulting state of the ensemble is termed a  conditional spin-squeezed state, the reduction in noise is completely deterministic with no discarding of trials necessary.  The improvement in phase estimation relative to the SQL is quantified by the observed spectroscopic enhancement\cite{Wineland1992} $W^{-1} \equiv \left(\Delta \theta_{SQL} / \Delta \theta \right)^2$. $W^{-1}>1$ also guarantees the state is entangled. 

First proof-of-principle experiments have generated conditional spin-squeezed states by using collective pre-measurements\cite{AWO09,SLV10,Wasilewski2010,CBS11,SKN12}. However, measurement-induced back-action, including loss of sensitivity due to decoherence and noise due to population diffusion between states, has limited direct observations of spectroscopic enhancement to $W^{-1} \le 1.4$ \cite{Wasilewski2010}.
Such modest improvements have cast doubt on the effectiveness of using pre-measurements for entanglement-enhanced metrology. 

In this Report, we realize collective pre-measurements with reduced back-action to achieve a directly-observed spectroscopic enhancement $W^{-1} = 10.5(1.5)$ in an ensemble of $N=4.8\times 10^5$  $^{87}$Rb atoms.  
We emphasize that this result reflects no background subtraction or corrections for finite probe detection efficiency, which is critical for the realization of practical applications of entangled states.  

Entanglement-enhanced states have also been generated for small ($N \le 14$) collections of atomic ions using quantum-logic operations, and for much larger neutral atomic ensembles using atomic collisions or cavity-mediated optical feedback.  Directly observed enhancements of up to $W^{-1} = 5.0(1)$, $7(1)$, and $3.6(5)$ (at $N=8$, $4.5\times 10^4$, and $3\times 10^4$) have been realized using these respective approaches\cite{Monz2011,Chapman12,LSM10}, with additional references listed in supplementary text.  Squeezed microwave\cite{Eichler2011} and optical\cite{Eberle2010} fields have achieved equivalent enhancements of up to $1.6$ and $18.6$ respectively.

Our experimental system consists of an ensemble of $N$ pseudo-spin-1/2s formed by the hyperfine ground states $\ket{\uparrow} \equiv \ket{F=2, m_f=+2}$  and $ \ket{\downarrow} \equiv \ket{F=1, m_f=+1}$ in $^{87}$Rb, separated by 6.8~GHz. The quantum state of the ensemble can be approximated as a single collective spin or Bloch vector ${\bf J\equiv\langle\hat{J} \rangle}$, in an abstract space defined by the collective spin operator $\hat{{\bf J}} =  \hat{J}_x \hat{x}+ \hat{J}_y \hat{y} + \hat{J}_z\hat{z}$  (Fig. 1B and \cite{OSM}). The spin projection operator $\hat{J}_z = \hat{N}_\uparrow-N/2$ can be written in terms of the measurable quantities:  total atom number $N$ and the spin up population operator $\hat{N}_\uparrow = \Sigma_{i=1}^N \ket{\uparrow_i} \bra{\uparrow_i}$, where $i$ labels individual atoms. The length of the vector is $J = \left| \left< \hat{{\bf J}} \right> \right|$. For an unentangled CSS, $J=N/2$. 

The quantum projection noise and standard quantum limit can be understood as arising from uncertainty in the orientation of the Bloch vector (Fig. 1B).  This quantum uncertainty can be visualized as a quasi-probability distribution perpendicular to the mean vector.  When the Bloch vector is oriented along $\hat{x}$, the degree of uncertainty in the orthogonal spin projections is constrained by a Heisenberg uncertainty relationship $\Delta J_z \Delta J_y\ge N/4$, where $\Delta X$ indicates the standard deviation of repeated measurements of $X$.  For a CSS of atoms,  $\Delta J_z= \Delta J_y= \Delta N_{CSS} = \sqrt{N}/2$. For the the polar angle $\theta \approx J_z/J = 2N_\uparrow/N - 1$ measured from the equator of the Bloch sphere, the SQL is set by the projection noise fluctuations to $\Delta \theta_{SQL} = 1/\sqrt{N}$. 

The conditionally squeezed state is created by first preparing a CSS along $\hat{x}$ and then making a collective pre-measurement $\hat{N}_\uparrow$, with measurement outcome labeled $N_{\uparrow p}$, and subtracting the result from a subsequent final measurement $\hat{N}_\uparrow$, labeled $N_{\uparrow f}$. The differential quantity $N_{\uparrow f}-N_{\uparrow p}$ can possess reduced noise relative to the projection noise fluctuations $\Delta N_{CSS}$ appearing in the two separate measurements. The spin noise reduction is calculated as $R=(\Delta(N_{\uparrow f}~-~N_{\uparrow p}))^2 / \Delta N_{CSS}^2$. By making a collective or joint measurement, any rotation of the vector's polar angle $\theta$ that occurs between the two measurements will still modify the differential quantity $N_{\uparrow f}-N_{\uparrow p}$, leading to the desired enhancement in the estimation of applied phase shifts.

To measure the collective state population $N_\uparrow$, the atomic ensemble is coupled to the TEM$_{0,0}$ mode of an optical cavity. The coupling is characterized by an effective single-atom coupling $g = 2\pi\times 450(20)$ MHz. The details of inhomogeneous coupling to the probe in our standing wave cavity are handled as in refs. \cite{SLV10,CBS11}. With no atoms present, the cavity has a resonant frequency $\omega_c$ and decay rate $\kappa = 2\pi \times 11.8(1)$ MHz. We detune the cavity frequency from an atomic transition by $\delta = \omega_c-\omega_a = 2\pi \times 200$ MHz, where $\omega_a$ is the frequency of the atomic transition from $\ket{\uparrow}$ to an optically excited state $\ket{e} \equiv \ket{F=3', m_f=+3}$. The radiative decay rate of $\ket{e}$ in free space is $\Gamma = 2\pi \times 6.07$ MHz. Atoms in $\ket{\uparrow}$ produce a dressed atom-cavity resonance at frequency $\omega_{c'}$, such that $\omega_{c'} - \omega_c = (\sqrt{\delta^2 + 4g^2N_\uparrow} -\delta)/2$ (Fig. 1C and \cite{OSM}).  We measure $\omega_{c'}$ with a probe laser (frequency $\omega_p$) to determine $N_\uparrow$. The strength of the collective measurement is characterized by the average number of probe photons $M_t$ transmitted through the cavity. The probe provides collective population information corresponding to the total number of atoms in $\ket{\uparrow}$, without providing individual atomic state information.

The information gained from a pre-measurement $\hat{N}_\uparrow$ causes back-action on the system, illustrated in Fig. 1D.  First, the measurement reduces the collective spin projection uncertainty to $\Delta J_z = \Delta N_{\uparrow m}$, where $ \Delta N_{\uparrow m}$ is the the measurement imprecision.  The Heisenberg uncertainty relationship requires fundamental back-action to appear in the orthogonal spin projection $\Delta J_y \ge (N/4)/\Delta N_{\uparrow m}$, referred to as anti-squeezing. Because $J_z$ is not coupled to the back-action quadrature $J_y$, the ideal measurement is intrinsically back-action evading\cite{Braginsky1980}.  

However, real systems experience at least two additional sources of probe-induced back-action, also illustrated in Fig. 1D. Both are caused by photons spontaneously scattered from the probe into free space, with average number of scattered photons $M_s$ scaling linearly with the measurement strength $M_s \propto M_t$. One source of back-action arises from free-space scattered photons leaking single-atom information to the environment, projecting an individual atom into $\ket{\uparrow}$ or $\ket{\downarrow}$ for every free-space scattered photon. The result is a shortening of the Bloch vector such that a subsequent angular deflection $\theta$ will produce a reduced change of the measured population $N_{\uparrow f}$. 

Another source of probe-induced back-action is spontaneous transitions between ground states driven by the same free-space scattering.  Quantum randomness in the number of transitions between states adds noise to the measurement of $N_\uparrow$ as the population diffuses amongst ground states.  The added noise $\Delta  N_{\uparrow D}$ scales as $(\Delta  N_{\uparrow D})^2\propto p M_t$, where $p$ is the probability an atom changes state if it scatters a photon into free-space. The optimum spin-noise reduction $R$ is fundamentally limited by the need to balance the decrease in measurement noise $\Delta N_{\uparrow m} \propto 1/\sqrt{M_t}$ versus the increase in diffusion noise $\Delta  N_{\uparrow D} \propto \sqrt{M_t}$. This balancing is analogous to radiation pressure back-action that sets the SQL for measurements of mechanical position\cite{Braginsky1980}.

The key experimental advance in this work is the elimination of state-changing transitions as a limitation on $W$. This is achieved by creating a system in which collective coupling to the probe mode is enhanced relative to single-atom processes. Our approach uses the medium finesse optical cavity $F=660$ to enhance the collective coupling, with the figure of merit $N C\approx 6\times 10^3$, where $C = 1.1(1)\times10^{-2}$ is the single-atom cooperativity \cite{SLV10,CBS11}.  In addition, we suppress state-changing transitions by using $\sigma^+$ polarized probe light on a cycling transition \cite{Saffman2009,AWO09,Zhang2012,CBW13}. If a photon is scattered into free space, the ground-state transition probability $p$ is $\sim 1/150$ that of our previous work\cite{CBS11}. As a result, previously ignored noise sources, not yet fully understood (see supplementary text), now dominate the probe-induced back-action on the measurement $\hat{N}_\uparrow$.

In Fig. 2, we directly sense an externally-applied phase shift with resolution below the SQL. 
We apply a small rotation $\psi$ of the polar angle $\theta$ using a microwave pulse,  described in Methods \cite{OSM}. 
In one case, the rotation is applied to a CSS with no pre-measurement. In a second case, the rotation is applied just after the pre-measurement of $N_\uparrow$ prepares a conditional spin-squeezed state.
The deflection of $N_{\uparrow f}$ is slightly smaller for the spin-squeezed state due to probe-induced collapse during the pre-measurement $N_{\uparrow p}$.  
However, the reduction in noise in the quantity $N_{\uparrow f} - N_{\uparrow p}$ allows the rotation angle to be estimated with an enhancement $W=7.5(9)$ in this example data set with measurement strength $M_t = 2.7(1)\times10^4$ and $N = 4.3\times10^5$.  No background subtractions or corrections are applied.  In a single-shot, the fractional error rate in determining whether the phase shift of $\psi=2.3(1)$ mrad was applied is reduced from 0.27(1) without the pre-measurement to 0.022(7) with the pre-measurement. 

More generally, we can identify an optimum spectroscopic enhancement by measuring both the spin noise reduction $R$ and the fractional shortening of the Bloch vector $\mathcal{C} = J/(N/2)$ as a function of measurement strength $M_t$, shown in Fig 3A. First we consider the spin noise reduction.
The maximum $R^{-1}$ observed, with no background subtraction, is $R^{-1}= 16(2)$ at $M_t = 4.1\times 10^4$. 
The contributions of various noise sources are quantified using a fit to the observed $R$ versus $M_t$.  
The model,  $R = r_{PSN}/M_t+R_{tf}+r_q M_t+r_c M_t^2$, includes four noise contributions:  photon shot noise $r_{PSN}$, a technical noise floor $R_{tf}^{-1} =73(34)$ independent of $M_t$,  probe-induced quantum back-action $r_q$, and probe-induced classical back-action $r_c$.  
Photon shot noise dominates at low $M_t$ so that $R^{-1}$ initially increases as $M_t$ increases. However, the rise in classical back-action $r_c M_t^2$ eventually limits $R^{-1}$. At the optimum $M_t$, the classical back-action $r_c$ alone would limit $R^{-1}$ to $67(15)$. The quantum back-action $r_q$ is statistically consistent with zero.

We have effectively eliminated ground state transitions as a substantial source of back-action in the current experiment. The noise added due to population diffusion from state-changing transitions is estimated to only limit $R^{-1}$ to $1.7(3)\times10^3$ as measured by probe-induced optical pumping between different ground-states (see supplementary text). Also, the inferred contribution to $r_c$ due to the observed classical fluctuations in probe power would only limit $R^{-1}$ to $3.2(4)\times10^{4}$.  The equivalent transition probability is $p \le 4.4(8)\times10^{-3}$. For comparison, the clock states in our previous work had a transition probability $p=2/3$ which, in our current system, would limit $R^{-1}$ to $1.9(2)$. With population noise considerably suppressed, other sources of back-action, including optomechanical effects (Fig. 3B), appear to dominate the probe induced back-action on $R$.

Free-space scattering also leads to a reduction in the Bloch vector length $J$, and the resulting loss of signal must be accounted for to determine the spectroscopic enhancement. To determine $J$, the polar angle of the Bloch vector $\theta$ is varied after the pre-measurement $N_{\uparrow p}$ using a microwave pulse. The population $N_\uparrow$ is then recorded versus the rotation angle, shown in Fig. 3C.  The fractional reduction in length of the Bloch vector is determined from the fitted contrast $\mathcal{C} = 2 J/N$ of the observed fringe. The initial contrast at $M_t =0$ is $\mathcal{C}_i=0.97(3)$, and $\mathcal{C}$ monotonically decreases as a function of $M_t$, close to the limit from wave function collapse due to free space scattering (Fig. 3A).  We believe uncanceled inhomogeneous probe light shifts are responsible for the additional small loss of contrast (see supplementary text).

Taken together, the decrease in spin noise and loss of contrast quantify the spectroscopic enhancement of the spin-squeezed state $W^{-1}  = R^{-1} \mathcal{C}^2/\mathcal{C}_i$, as calculated in refs. \cite{AWO09,SLV10,CBS11}.  
The optimum observed improvement corresponds to $W^{-1} =10.5(1.5)$ or $10.2(6)$~dB. This value includes no measurement background noise subtraction, and thus represents the actual realized improvement in phase sensitivity. 

Further confirmation that our collective measurement is near the cycling transition limit is the observed $\Delta \theta^2 \propto 1/N^{2}$ scaling of the absolute phase resolution\cite{AWO09,CBW13}, shown in Fig. 4. 
For comparison, the optimal phase resolution when state-changing processes are the dominant limitation on $W$ scales as $N^{-3/2}$. This more favorable scaling with $N$ is important for practical applications where absolute phase resolution is the figure of merit.

When the spectroscopic enhancement $W^{-1} \ge 1$, the ensemble is guaranteed to be entangled.  Maximally entangled ensembles can achieve phase estimation precision of $\Delta \theta_{HL}^2= 1/N^2$, known as the Heisenberg limit, which has been realized with small ensembles\cite{Monz2011}.  While our system is far from the Heisenberg limit for $N = 4.8\times 10^5$ atoms, the absolute phase sensitivity is equivalent to $\sim 44000$ copies of a maximally-entangled $11$ atom ensemble.  Such a comparison emphasizes the massive parallelism achievable by collective measurements to generate entanglement in neutral atom ensembles.


The optical nature of our approach, among others\cite{AWO09,SLV10,Wasilewski2010,LSM10,SKN12}, offers the advantage that the probe or squeezing laser can be completely extinguished after the squeezed state is generated. In contrast, a potential challenge for squeezing generated using atomic collisions is whether the interactions that generate entanglement can be sufficiently reduced to avoid loss of accuracy and precision during the subsequent sensing period. Our approach is compatible with a wide array of atomic sensors but is particularly appealing for optical lattice clocks\cite{Hinkley2013}, where systematic errors and atom loss may place a limit on the ensemble size. The probe also provides a low-noise, non-destructive readout method useful for purely classical, but substantial, improvements in optical lattice clocks and other atomic sensors \cite{Westergaard2010}.  

Straightforward technical improvements could both decrease the technical measurement noise floor and increase the total probe detection efficiency from $8(5)\%$ to $>50\%$, allowing us to reach $W^{-1} \sim 100$ in our current system with only a medium finesse optical cavity.  However, to realize even further spectroscopic enhancement, previously unimportant forms of probe-induced back-action will likely require additional study. 


\paragraph*{Acknowledgements:} The authors would like to acknowledge Katherine McAlpine's early contributions to building detectors and helpful discussions with Ana Maria Rey and Konrad Lehnert. All authors acknowledge financial support from DARPA QuASAR, ARO, NSF PFC, and NIST.  J.G.B. acknowledges support from NSF GRF, K.C.C. acknowledges support from NDSEG, and Z.C. acknowledges support from A*STAR Singapore. This material is based upon work supported by the National Science Foundation under Grant Number 1125844. Correspondence should be addressed to J. G.B. (bohnet@jilau1.colorado.edu) or J. K. T. (jkt@jila.colorado.edu).



\clearpage

\begin{figure}
\includegraphics[width = 6.2in]{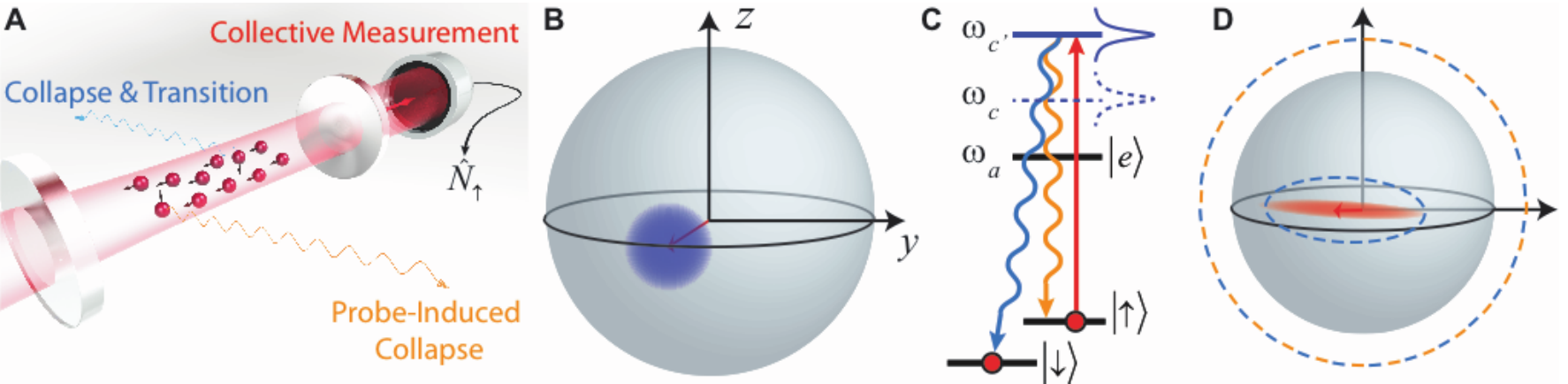}
\caption{{\bf Spin-squeezing and measurement back-action} ({\bf A}) Atoms collectively interact with light in an optical cavity. A measurement of the phase of the probe field (red) is sensitive to the total number of atoms in spin up, and projects the ensemble into an entangled state, conditioned on the measurement outcome. Probe photons can be scattered into free space, causing atoms to collapse to spin up (orange in A, C, and D) and can also cause state-changing transitions (blue in A, C, and D). ({\bf B}) A coherent spin state can be visualized by a Bloch vector (red), with a pointing uncertainty set by quantum noise, represented by the shaded uncertainty disk. ({\bf C}) Atoms in $\ket{\uparrow}$ with optical transition frequency $\omega_a$ couple to the detuned cavity mode with resonance frequency $\omega_c$. The coupling results in a dressed cavity mode with resonant frequency $\omega_{c'}$, so probing $\omega_{c'}$ measures the total number of atoms in $\ket{\uparrow}$, and hence the Bloch vector's spin projection $J_z$, without measuring the state of individual atoms. 
Probing on a cycling transition suppresses back-action from scattering events that change an atom's state to $\ket{\downarrow}$ (blue), limiting back-action to collapse (orange). ({\bf D}) After a pre-measurement, back-action modifies the noise distribution on the Bloch sphere. Fundamental back-action appears along $\hat{y}$. Back-action from non-ideal measurements, indicated by dashed lines, include reduction in length $J$ of the collective Bloch vector and added noise in $J_z$ caused by state-changing transitions.  
}
\end{figure}

\begin{figure}
\includegraphics[width=6.2in]{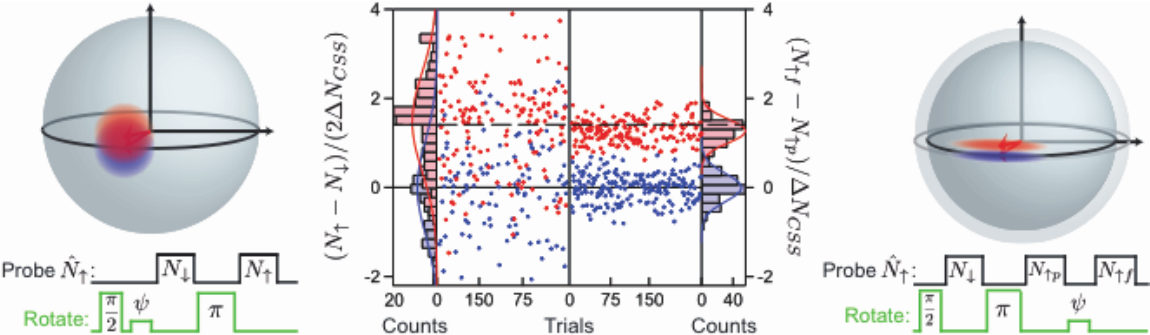}
\caption{{\bf Detection of a quantum phase with entanglement-enhanced sensitivity.} 
We apply a small rotation $\psi$ to the polar angle $\theta$ of both a CSS and a spin-squeezed state, with data and representative Bloch spheres shown on the left and right sides respectively. Red data points show experimental trials with $\psi = 2.3(1)$ mrad, and blue data points show trials with $\psi = 0$. The data are represented both as histograms and Gaussian curves generated from the average and standard deviation of the measurements.  The experimental timing sequence consists of probe pulses (black) and microwave rotation pulses (green). For the CSS, the rotation $\psi$ is applied immediately after preparing the CSS along $\hat{x}$. The rotation $\psi$ appears as a change in the quantity $N_\uparrow-N_\downarrow$, which is normalized to the total projection noise that appears in this differential quantity. In the case of the spin-squeezed state, we perform the rotation $\psi$ after a pre-measurement $N_{\uparrow p}$. The rotation then appears as a change in $N_{\uparrow f}-N_{\uparrow p}$, where the projection noise largely cancels. The spin-squeezed state has a precision $W^{-1} = 7.5(9)$, even though the change in $N_{\uparrow f}-N_{\uparrow p}$ is slightly smaller than in the CSS 
due to free-space scattering during the pre-measurement.  The loss of signal is represented by a smaller Bloch sphere for the spin-squeezed state.}
\end{figure}

\begin{figure}
\includegraphics[width = 6in]{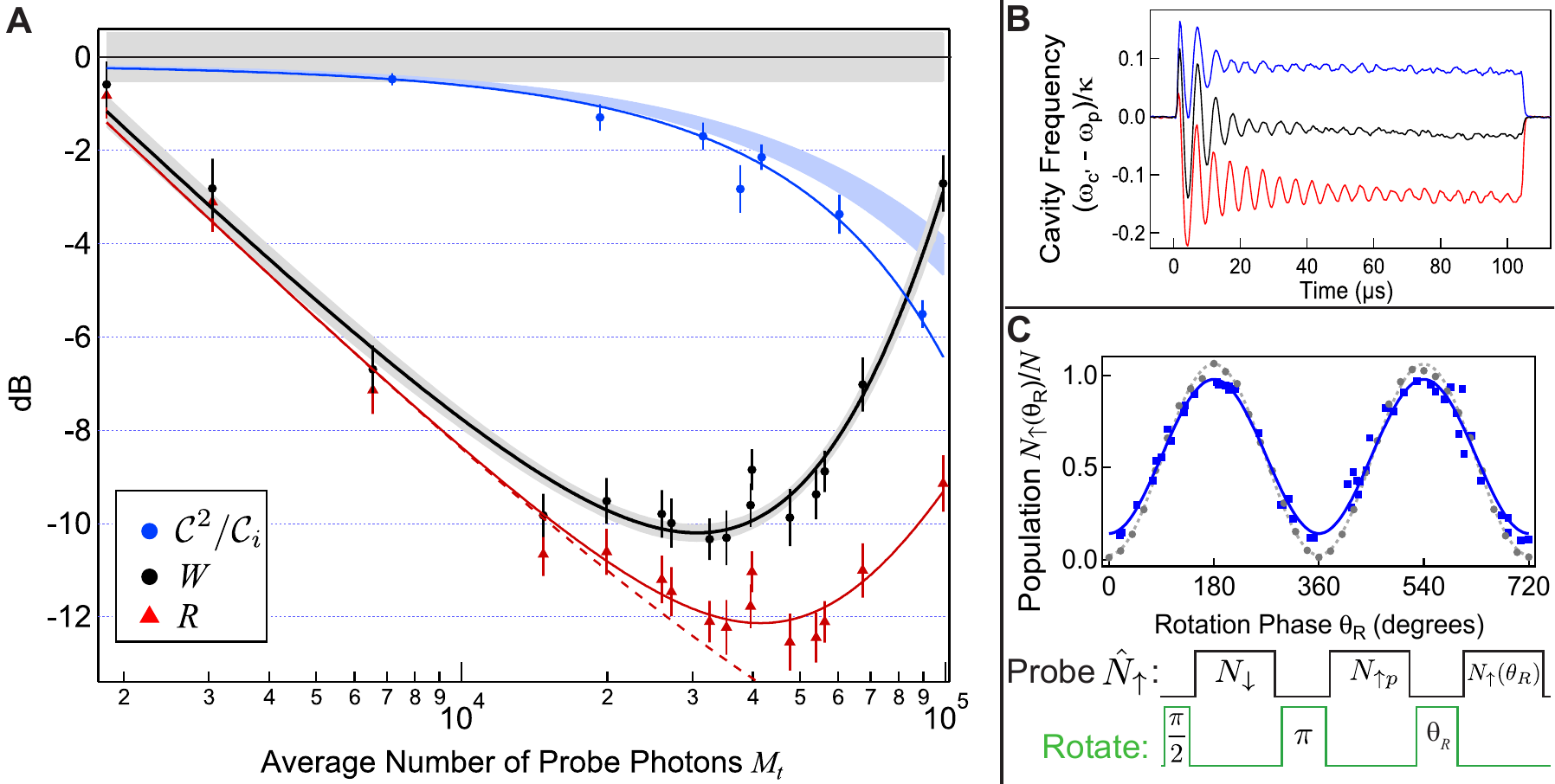}
\caption{{\bf Spin-squeezing and probe-induced back-action.} ({\bf A}) Scaling of the spin noise reduction $R$ (red), loss of signal $\mathcal{C}^2/\mathcal{C}_i$ (blue), and the inverse of the spectroscopic enhancement $W$ (black) versus probe intensity $M_t$ for $N = 4.8\times10^{5}$. 
The red, blue, and black curves are fits to the data. The data for $W$ is calculated from $R$ data and the fit to $\mathcal{C}^2/\mathcal{C}_i$. The $68\%$ confidence band for the $W$ fit and the SQL is in grey. The dashed red curve shows the fitted $R$ assuming no probe-induced added noise ($r_c = r_q = 0$). The light-blue region is the predicted $\mathcal{C}^2/\mathcal{C}_i$ due to free space scattering. All error bars are 1 std.\ dev. We use the usual convention for expressing a ratio $X$ in dB units, $x$ (dB) $ = 10\log_{10}X$.  ({\bf B}) Examples of optomechanical oscillations in the dressed cavity frequency $\omega_{c'}$. The relative detuning of $\omega_{c'}$ and $\omega_p$ results in increased or decreased oscillation damping rates, a source of probe-induced back-action noise (see supplementary text). Each curve is the average of 30 experimental trials. 
({\bf C}) Example data and experimental sequence for the measurement of the contrast $\mathcal{C}$. Probe pulses (black) are measurements  $\hat{N}_\uparrow$. Microwave pulses (green) rotate the polar angle $\theta$ of the Bloch vector. After the pre-measurement of $N_{\uparrow p}$, a variable rotation $\theta_R$ is applied and $N_\uparrow(\theta_R)$ is recorded. The contrast $\mathcal{C}$ is determined from the amplitude of the $N_\uparrow(\theta_R)$ fringe (curves are a fit to the data), with two examples shown in blue and grey for $M_t = 3.0\times10^4$ and $M_t = 0$ respectively.}
\end{figure}

\begin{figure}
\centering
\includegraphics[width=4.5in]{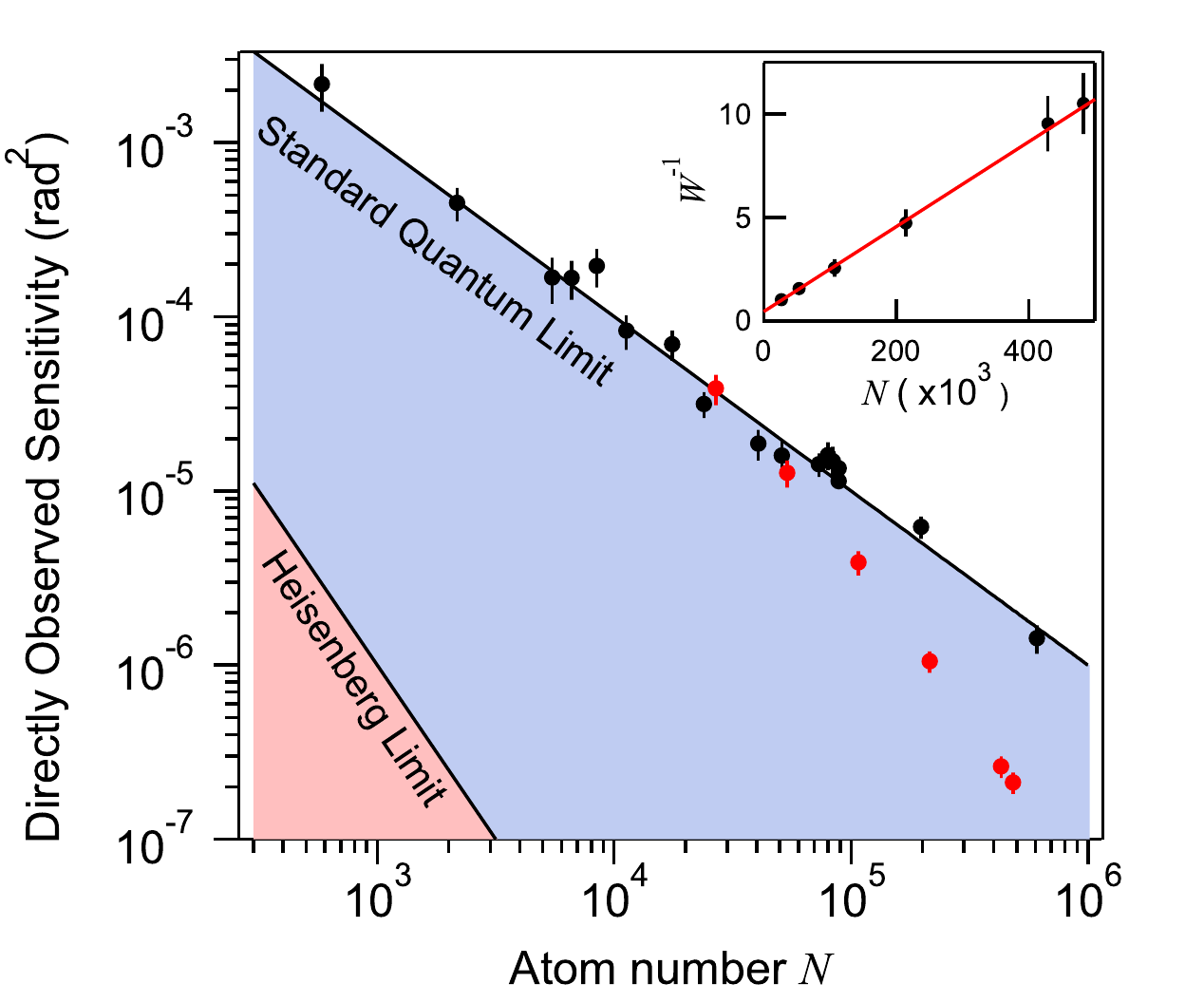}
\caption{{\bf Absolute phase sensitivity versus \textit{N}.} The red points are the observed spin-squeezed state phase sensitivities (optimized with respect to measurement strength $M_t$) for different atom numbers $N$. The data show the predicted $1/N^2$ scaling for probing on a cycling transition \cite{Saffman2009,CBW13}, equivalent to a linear scaling of the spectroscopic enhancement $W^{-1}$ versus $N$, shown in the inset. The red line is a linear fit to the data. The SQL is confirmed by measuring the  projection noise that appears in $N_{\downarrow} - N_{\uparrow p}$ (black points, each 100 trials) and observing $1/\sqrt{N}$ scaling. Error bars indicate $68.3\%$ confidence intervals.}
\end{figure}


\clearpage


\renewcommand\thefigure{S\arabic{figure}}    
\setcounter{figure}{0} 

\renewcommand\thetable{S\arabic{table}}    
\setcounter{table}{0} 

\section*{Supplementary Material}

\subsection*{Supplementary references for entanglement-enhanced states}
Generating ensembles with angular precision enhanced by entanglement is a very active and rapidly advancing field.  For a more complete list of other experimental realizations of entanglement-enhanced states in atomic ions, see refs. \cite{Wineland01,Leibfried2005,Monz2011,Noguchi2012} and in neutral atomic ensembles, see refs. \cite{EGW08,RBL10,GZN10,LSM10, Bucker2011,Klempt11,Chapman12}.

\subsection*{Materials and Methods}

\subsubsection*{Atom-cavity system}
The optical cavity used for the collective measurements has a measured free spectral range of $7.828(1)$ GHz and a measured transverse mode spacing of $2.257(2)$ GHz, determining the cavity length to be $L = 1.9149(3)$ cm and the mode waist to be $w_{780} = 69.90(4)$ $\mu$m at 780 nm. With no atoms in the cavity, the cavity frequency is denoted $\omega_c$ and the measured cavity linewidth or power decay rate is $\kappa = 2\pi \times 11.8(1)$ MHz, giving a cavity finesse $F=663(5)$. The power decay rate from factory-specified mirror transmission alone is $\kappa_\circ = 2\pi \times 5.02$ MHz.

The cavity is also used to generate the one-dimensional, intra-cavity optical lattice trap at wavelength $\lambda_l=823$~nm, with a mode waist $w_{823} = 71.78(4)$ $\mu$m and an axial trap frequency $\omega_{ax} = 2 \pi \times 150$ kHz.
The atoms are polarization-gradient-cooled to 10(2)~$\mu$K in the trap, putting them in the Lamb-Dicke regime along the the cavity axis, but not in the transverse direction.
A magnetic field with magnitude $|{\bf B}| = 0.73(1)$ G is oriented along the cavity axis.  A simplified diagram of the experimental setup is shown in Fig. S1.

The atomic ensemble is prepared and probed using the D2 line in $^{87}$Rb at 780 nm (Fig S2). The hyperfine ground states $\ket{\downarrow}\equiv \ket{5~^{2}S_{1/2}, F=1, m_f = 1}$ and $\ket{\uparrow} \equiv \ket{5~^{2}S_{1/2},F=2, m_f=2}$ form the pseudo-spin-1/2 system. Coherent rotations of the spin system are accomplished by coupling the ground states with a 6.833~GHz microwave field\cite{CBW12}. The relevant optically excited state $\ket{e} \equiv \ket{5~^2P_{3/2}, F' = 3, m_f=3}$ has a decay rate $\Gamma = 2\pi\times 6.06$ MHz. 


The coupling between the cavity mode and a single atom on the $\ket{\uparrow}$ to $\ket{e}$ transition is parametrized by the single-photon Rabi frequency $2 g_0 = 2\pi \times 1070(30)$ kHz, given at an anti-node of the standing wave probe field at the center of the cavity both in the transverse and axial dimensions\cite{Kimble1998}. The probe coupling varies sinusoidally as $g(z) = g_0 \cos{\left(2\pi z/\lambda_p\right)}$, where $z$ is the spatial coordinate along the axis of the cavity. The probe field at wavelength $\lambda_p=780$~nm is incommensurate with the trapping site spacing set by $\lambda_l$, so we define an effective single photon Rabi frequency $2g$  for the ensemble, accounting for both the axially and radially varying coupling as in \cite{SLV10,CBS11}: $2g =2\pi \times 894(46)$ kHz (or cooperativity parameter $C = 4g^2/(\kappa\Gamma) = 1.1(1)\times10^{-2}$). The total number of atoms confined in the optical lattice is $N_\circ$, irrespective of their coupling to the standing-wave probe mode.  An effective atom number $N = 0.663(4) \times N_\circ$ with uniform coupling $2g$ produces the observed projection noise fluctuations.  Throughout the main text, $N$ refers to the effective atom number.



\subsubsection*{Collective population measurement}
The collective measurement of atomic population in $\ket{\uparrow}$, $\hat{N}_\uparrow$, is made by measuring the resonance frequency of the coupled, or dressed, atom-cavity system $\omega_{c'}$. The collective coupling of the atoms in $\ket{\uparrow}$ to the cavity mode is $\Omega_\uparrow = \sqrt{N_\uparrow}2g$, where $N_\uparrow$ is number of atoms in $\ket{\uparrow}$. The bare cavity resonance frequency $\omega_c$ is detuned $\delta = \omega_c - \omega_a = 2 \pi \times 200$~MHz to the blue of the $\ket{\uparrow}\rightarrow\ket{e}$ transition at frequency $\omega_a$.  The collective coupling results in a dressed cavity resonance frequency that is shifted by an amount

\begin{equation}
\omega_{c'} - \omega_c =\frac{1}{2}\left( \sqrt{\delta^2+ N_\uparrow \left (2 g\right )^2}-\delta \right) \,\,.
\label{eqn:dressscavityfrequency}
\end{equation}


\noindent We measure the frequency of the dressed atom-cavity system using $\sigma^+$ polarized light from a narrow ($<$~5 kHz) linewidth probe laser similar to that in ref. \cite{Lin2012}.  The probe frequency $\omega_p$ is tuned nominally to resonance with the dressed cavity mode $\omega_{c'}$. The phases of the transmitted and reflected probe signals give the resonance frequency $\omega_{c'}$.
The measured frequency is then used in Eqn. S1 to determine $N_\uparrow$. The relevant cavity and probe frequencies are schematically represented in Fig. 2S.

The collective measurement strength is quantified by $M_t$, the average number of photons transmitted through the cavity in a single measurement $\hat{N}_\uparrow$. For each measurement, the probe light is on for 43 $\mu$s, but the measurement is an average of $\omega_{c'}$ over only 40 $\mu$s to avoid edge effects in the data acquisition. Probe-induced back-action from free space scattering is quantified by $M_s$, the number of photons scattered into free space during each measurement. $M_s$ is related to $M_t$ by $M_s = M_t \left( \frac{2\Gamma}{\kappa_\circ}\frac{N_\uparrow 4g^2}{4 (\omega_{c'}-\omega_a)^2} \right)$ \cite{CBW13}.  For the data with optimum squeezing at $N = 2N_\uparrow = 4.8\times 10^5$, the calculated number of photons scattered into free space is $M_s = 1.0(1) \times M_t$.

\subsubsection*{Spin noise reduction measurement sequence}
Spin noise reduction $R$ is determined from the directly measured noise in the difference of two measurements of the population $N_\uparrow$. The experimental sequence for measuring $R$ is shown in Fig. S3. We first pre-align the probe frequency $\omega_p$ to the dressed cavity frequency $\omega_{c'}$, described in detail below. Next, we prepare a coherent spin-state on the equator of the Bloch sphere by optically pumping all atoms to $\ket{\downarrow}$ ($\bf{J}$ along $-\hat{z}$), then applying a $\pi/2$ microwave pulse to orient the Bloch vector along $\hat{x}$, equivalent to placing each atom in a superposition of $\ket{\uparrow}$ and $\ket{\downarrow}$.  
We infer the number of atoms in $\ket{\uparrow}$ by switching on the probe laser to make a measurement of $\omega_{c'}$, and label the result $N_\downarrow$.  
After phase-coherently swapping the populations in $\ket{\uparrow}$ and $\ket{\downarrow}$ using a microwave $\pi$-pulse, we perform two successive measurements of the population in $\ket{\uparrow}$ and label the results $N_{\uparrow p}$ and $N_{\uparrow f}$.  
The experiment is repeated  more than 100 times and $R$ is computed from the variance of the difference between the final measurement and the pre-measurement $(\Delta\left(N_{\uparrow f}-N_{\uparrow p}\right))^2$, as $R= \frac{\left(\Delta\left(N_{\uparrow f}-N_{\uparrow p}\right)\right)^2}{N/4}$. Throughout the main text and supplementary material, $\Delta X$ is used to indicate the standard deviation of repeated measurements of the quantity $X$. 

In Fig. S4, we compare measurement noise versus $M_t$ both with and without atoms in the cavity. With atoms, the contribution from photon shot noise is 4.5 dB lower because the presence of atoms creates a narrower dressed cavity resonance \cite{CBW13}. The difference is visible for small $M_t$ where photon shot noise dominates. However, measurement back-action with atoms present contributes noise that increases with $M_t$ and limits the maximum-achievable spin noise reduction $R$, as described later in the supplementary text. 


The role of the first measurement, labeled $N_\downarrow$ in Fig. S3, is to cancel inhomogeneous light shifts caused by the standing-wave probe light during the pre-measurement $N_{\uparrow p}$, with the $\pi$-pulse forming a spin-echo sequence. Light shift cancellation is critical for restoring coherence (i.e. length of the Bloch vector) so that a Bloch vector rotation through an angle $\theta$ after the pre-measurement will produce an observable change of the final measurement $N_{\uparrow f}$.

In principle, the first measurement can be used to enhance the spin noise reduction. However, we found that the $\pi$-pulse degraded the observed spin noise reduction when the information gained from the first measurement was utilized. The noise added by the microwave rotation results from both fluctuations in the transition frequency \ket{\downarrow} to \ket{\uparrow}, as well as amplitude and phase noise of the applied microwave field.  Neglecting the information in $N_\downarrow$ avoids these sources of rotation-added noise at the cost of a potential factor of 2 improvement in $R$.

Fluctuations in the atom number $N$ produce significant fluctuations in the detuning $\delta_p = \omega_p - \omega_{c'}$ of the probe from the dressed cavity resonance from one trial to the next.  For scale, the atom number changes by $\sim1\%$ peak to peak on 1 minute time scales, and can drift by $2.5\%$ over 30 minutes.  Fluctuations in $\delta_p$ can create additional technical noise that limits $R$.  First, the number of transmitted photons $M_t$ changes with the detuning of the probe, modifying the small observed chirping of $\omega_{c'}$ described below.  Secondly, the point of maximum measurement sensitivity is achieved when $\delta_p \ll \kappa/2$.

To mitigate these two effects, we implement a scheme to pre-align the probe laser frequency $\omega_p$ to the dressed cavity frequency $\omega_{c'}$ at the start of each trial (Fig. S3). After we prepare a coherent spin-state on the equator of the Bloch sphere by optical pumping and a $\pi/2$ rotation, we actively lock $\omega_p$ to $\omega_{c'}$, then hold $\omega_p$ fixed at the final frequency for all subsequent measurements within a trial.  After this pre-alignment, the ensemble is optically repumped to $\ket{\downarrow}$ and the measurement sequence described previously is performed. Pre-alignment of the probe laser reduced the trial to trial standard deviation of the probe-cavity detuning during the pre-measurement $N_{\uparrow p}$ to $\Delta \delta_p = 0.045 \times \kappa/2$.  For comparison, the fundamental limit set by the uncorrelated projection noise appearing in both the pre-alignment and the pre-measurement is $\Delta \delta_p = 0.034 \times \kappa/2$.





\subsubsection*{Collapse back-action}


To verify that the collective measurements generate a conditionally spin-squeezed state with enhanced phase sensitivity, we must determine the fractional reduction in the Bloch vector length $J$ caused by measurement back-action. The degree to which the measurement avoids collapse and dephasing is quantified by the contrast $\mathcal{C} = J/(N/2)$.  
To measure $\mathcal{C}$, we apply a variable polar angle rotation $\theta_R$ to the Bloch vector using an additional $\pi/2$ microwave pulse after the measurement of $N_{\uparrow p}$.
The polar angle is changed by varying the phase of the final microwave $\pi/2$ pulse relative to that of the initial $\pi/2$ pulse.  We then measure the population of $\ket{\uparrow}$ with measurement outcome labeled $N_\uparrow(\theta_R)$. The contrast is determined from fitting the data to its expected dependence $N_{\uparrow}(\theta_R) = (N/2)(1+\mathcal{C}\cos{\theta_R})$. 
Assuming each free space scattered photon causes a single atom to collapse into spin up, reducing $J$ by one unit, the predicted contrast is $\mathcal{C}_{pred} = \mathcal{C}_i e^{-M_s/N}$.  The initial contrast $\mathcal{C}_i = 0.97(3)$ is the contrast measured with no probe light ($M_s=M_t=0$) during the pre-measurement.  When $M_s$ is increased such that $\mathcal{C}$ falls below $\sim 0.9$, the contrast begins to decrease more rapidly than predicted by free space scattering, as seen in Fig. 3A.  We believe the deviation is likely due to light shifts that are not fully canceled by the spin-echo formed by the combination of measurements labeled $N_\downarrow$ and $N_{\uparrow p}$.

\subsection*{Ground state population back-action}
In this section, we quantify measurement back-action due to spontaneous scattering that changes the internal state of an atom.
We consider limitations to the spin noise reduction from both the fundamental quantum noise, which we call population diffusion, and noise from classical fluctuations in the average change in internal state populations.
For a perfect probing scheme on a cycling transition, only Rayleigh scattering is allowed, and so no change of the internal state of the atom would take place.  
However, as shown in Fig. S2 and discussed in ref. \cite{CBW13}, atoms in $\ket{\downarrow}$ can non-resonantly Raman scatter to both $\ket{\uparrow}$  and $\ket{1} \equiv \ket{F=2, m_f=1}$. Another possible measurement imperfection we consider is impure probe light polarization allowing population to Raman scatter out of $\ket{\uparrow}$.

Our model for the added noise in $R$ considers transitions between three states, $\ket{\uparrow}$, $\ket{\downarrow}$, and $\ket{1}$.  The probabilities per free space scattered photon to make a transition are denoted as $p_{\uparrow \downarrow}$, $p_{\downarrow \uparrow}$, $p_{\uparrow 1}$, and $p_{\downarrow 1}$ for the four transitions $\ket{\uparrow}\rightarrow\ket{\downarrow}$, $\ket{\downarrow}\rightarrow\ket{\uparrow}$, $\ket{\uparrow}\rightarrow\ket{1}$, and $\ket{\downarrow}\rightarrow\ket{1}$ respectively. To determine the transition probabilities, we do two experiments, shown in Fig. S5. In each experiment, we  measure an average change in $\omega_{c'}$ per transmitted probe photon. We then convert the frequency change into a transition probability using the calculated dressed cavity shift per atom added to each state $\alpha_s = \frac{d\omega_{c'}}{dN_s}$, where $s = \uparrow,\downarrow,1$.  In one experiment, shown in Fig. S5A, we measure the sum $p_{\downarrow \uparrow} + p_{\downarrow 1}$ and use known branching ratios to get the individual probabilities $p_{\downarrow \uparrow} = 7.3(7)\times10^{-4}$ and $p_{\downarrow 1} = 3.6(4)\times10^{-4}$. In the other experiment, shown in Fig. S5B, we directly measure $p_{\uparrow \downarrow} = 8(1)\times10^{-4}$. 
We assume $p_{\uparrow \downarrow}$ results from imperfect polarization of the probe laser. This measurement of $p_{\uparrow \downarrow}$ constrains the fraction of probe power in the non-cycling polarization to $< 6.7(8)\%$, if the imperfection is assumed to be $\sigma^-$-polarized light, and $< 1.7(2)\%$, assuming $\pi$-polarized light for the imperfection. Direct measurements more tightly constrain the fraction of power in the probe that is not in the $\sigma^+$ mode to $<0.5\%$. 
The final transition probability $p_{\uparrow 1}$ can be calculated using branching ratios, known probe detuning from atomic resonance $\omega_p-\omega_a$, and $p_{\uparrow \downarrow}$, assuming a specific ratio of $\sigma^-$ to $\pi$ polarization for the probe polarization imperfection.  For the full range of possibility for the polarization of the probe light imperfection (i.e. the ratio of $\sigma^-$ to $\pi$ polarized light in the probe imperfection ranging from 0 to 1), the transition probability doesn't change within its uncertainty, $p_{\uparrow 1} < 3.9(5)\times10^{-3}$. In all cases, these independent measurements show that the transition probabilities are small $p \ll 1$, approaching the ideal cycling transition limit of $p=0$.

The fundamental limitation on $R$ from internal state-changing scattering events comes from quantum noise in the scattering process. Specifically, if the average total number of transitions from $\ket{\uparrow}$ to $\ket{\downarrow}$ is $N_{\uparrow \downarrow} = p_{\uparrow \downarrow} M_s$, then on a given trial there will be quantum fluctuations in the total number of transitions with the standard deviation of $N_{\uparrow \downarrow}$ given by $\Delta N_{\uparrow \downarrow q} = \sqrt{p_{\uparrow \downarrow} \beta M_s}$. Here the factor of $\beta = 2/3$ accounts for the unweighted time averaging of the measurement records during the two measurements $\hat{N}_\uparrow$ that form the desired differential quantity $N_{\uparrow f} - N_{\uparrow p}$. In the limit that $M_s/(N/2)$ is small, the noise from each process is uncorrelated, and the noise added to $R$ is the sum of the individual variance of each process

\begin{align}
R_{pop,q} = \frac{\beta M_s}{N/4} \Big[ & p_{\uparrow \downarrow}(\alpha_{\downarrow} - \alpha_{\uparrow})^2 +p_{\uparrow 1}(\alpha_{1} - \alpha_{\uparrow})^2 \nonumber \\
&+ p_{\downarrow \uparrow}(\alpha_{\uparrow}-\alpha_{\downarrow})^2 +p_{\downarrow 1}(\alpha_{1}-\alpha_{\downarrow})^2 \Big]\, \, .
\end{align}
\noindent We assume no multiple Raman scattering, a good assumption for the low average probe photon number $M_t$ used here. 

Classical fluctuations in the probe photon number $M_t$, and hence $M_s$, also introduce classical noise in the number of Raman transitions between ground states which adds noise to $N_{\uparrow f} - N_{\uparrow p}$, limiting the spin noise reduction. Also, an average population change can add technical challenges to the experiment as the dressed cavity mode frequency $\omega_{c'}$ changes or chirps during a measurement. 
For our experiment, the classical rms fluctuations in $M_s$ are $\Delta M_{s} = 0.04 M_s$. These classical fluctuations contribute a term to the total spin noise reduction

\begin{align}
R_{pop,c} = \frac{\Delta M_{s c}^2}{N/4} \Big[& p_{\uparrow \downarrow}(\alpha_{\downarrow} - \alpha_{\uparrow}) +p_{\uparrow 1}(\alpha_{1} - \alpha_{\uparrow}) \nonumber \\
&+ p_{\downarrow \uparrow}(\alpha_{\uparrow}-\alpha_{\downarrow}) +p_{\downarrow 1}(\alpha_{1}-\alpha_{\downarrow}) \Big]^2 \, \, .
\end{align}

The noise added by classical probe power fluctuations exhibits a fortuitous cancellation in the current experiment. The added classical noise due to the population change is reduced by a factor of $\sim 6.5$ in variance due to a cancellation of the two terms containing $(\alpha_{\downarrow} - \alpha_{\uparrow})$ and as seen in Fig. S5.


The results of both the quantum and classical noise models, combined with the determination of the four transition probabilities, predict that internal-state changing transitions have been effectively eliminated as a source of probe induced back-action in this system (see Table S1 for constraints). Our measurements also indicate the added noise is small enough to allow for another two orders of magnitude improvement in spectroscopic enhancement for future experiments. 

\subsection*{Optomechanical back-action}

By suppressing internal-state transition noise, we reveal sources of probe-induced back-action that were not relevant in our prior measurement-induced spin-squeezing experiments \cite{CBS11}. Here we consider two sources that contribute noise to population measurements, both of which arise from the standing wave probe field in the cavity.
The first source of back-action results from dispersive forces exerted on the atoms by the probe, which cause coherent oscillations of $\omega_{c'}$ (previously observed in ref. \cite{Zhang2012}).  
The second source results from free-space scattering of probe photons leading to photon recoil heating, which causes a change of the dressed cavity frequency $\omega_{c'}$. 
In this section, we describe how these two effects contribute both classical and quantum noise to our measurement of $N_{\uparrow f} - N_{\uparrow p}$. Note that neither source of back-action acts as an additional source of decoherence (i.e. a loss of contrast $\mathcal{C}$). 
While these sources of back-action do not fully account for the measured back-action $r_c$ in our experiment, they are interesting as sources of back-action that may limit the spin noise reduction $R$ in future work. 


In our standing wave cavity, the probe induces a spatially varying light shift on the atoms, proportional to $g^2(z)$.  
The gradient of this light shift imparts a force on atoms not at a node or anti-node of the standing wave.  
When the probe is turned on, the atoms experience an impulsive momentum kick along the axial direction and oscillate at $\omega_{ax} = 2 \pi \times 150$ kHz in the confinement of the trapping lattice at 823 nm.  Because the coupling strength $g(z)$ is position dependent, the atomic motion causes $\omega_{c'}$ to oscillate at the same frequency.  The oscillations damp in approximately 10 $\mu$s due to a spread in axial oscillation frequencies caused by the spread in radial position of the atoms in the confining optical lattice. 


The oscillation is phase coherent between trials, and each measurement $\hat{N}_\uparrow$ largely averages over the oscillations. Nonetheless, variation in initial conditions lead to measurement noise in the differential quantity $N_{\uparrow f} - N_{\uparrow p}$ by causing variation in the collective oscillation. 
For example, classical and quantum fluctuations in $N_{\uparrow}$ cause variation in the probe detuning from the dressed cavity mode $\delta_p = \omega_p -\omega_{c'}$. 
The detuning $ \delta_p $ determines the nature of dynamic optomechanical effects on the atoms \cite{SchleierSmith2011}.
If $\omega_{c'} > \omega_p$, the collective oscillation experiences optomechanical damping. If $\omega_{c'} < \omega_p$, the oscillation will experience optomechanical anti-damping that lengthens the characteristic decay time of the observed oscillations of $\omega_{c'}$  (Fig. 3B). 


A change in the damping rate can alter the degree of cancellation of the optomechanical ringing in the differential measurement $N_{\uparrow f}-N_{\downarrow p}$. Pre-alignment of the probe frequency to the dressed cavity resonance, described previously, was important  for reducing fluctuations in the probe detuning $\delta_p$ that drive these variations.

We estimate the noise contribution of probe-induced collective oscillations by comparing data with and without the oscillations. We make the comparison by applying the probe continuously, and defining measurement windows that are time shifted with respect to the turn on of the probe where the ringing of $\omega_{c'}$ is largest.  We first calculate $\Delta (N_{\uparrow f} - N_{\uparrow p})$ measured using the first 86 $\mu$s of data, as was done in the spin-squeezing experiments. Then we calculate $\Delta (N_{\uparrow f} - N_{\uparrow p})$ using data starting 90 $\mu$s after the probe turn on when the collective oscillations have largely damped away. From the comparison of the two variances, we estimate that variable damping contributes a term $R_o$ to the spin noise reduction that empirically scales as $M_t^2$ and limits $R^{-1}$ to $620$ (Table S1). 


The second source of optomechanical back-action we consider here arises from free-space scattered photons heating the ensemble in a position-dependent manner.  Importantly, atoms that are more strongly coupled to the cavity mode are more strongly heated due to free-space scattering.  This process can be thought of as an external state-changing transition, paralleling the internal state-changing transitions discussed previously (i.e. Eqns. S2 and S3). The rise in temperature while probing changes the atoms' spatial distribution, reducing the overall coupling to the cavity and shifting $\omega_{c'}$. The average shift of $\omega_{c'}$ caused by the increase in temperature is 1.3 Hz per free space scattered photon, making it the largest average probe-induced frequency shift of $\omega_{c'}$ in our system. Thus, quantum and classical fluctuations in $M_s$ adds noise to measurements $\hat{N}_\uparrow$, contributing the relatively small quantum $R_{ext, q}$ and classical $R_{ext, c}$ terms to the spin noise reduction $R$ (see Table S1).

Again, the two identified sources of back-action, spatially-dependent photon-recoil heating and variable damping of the optomechanical ringing, fail to predict the entirety of the measured probe-induced back-action $r_c$ that scales as $M_t^2$ for data in Fig. 3A and S4. We believe the remaining added noise results from the details of the collective optomechanical oscillations, which are difficult to quantify and require further characterization. This additional noise will be important to understand for future advances but is beyond the scope of this work.

\clearpage

\begin{figure}
\centering
\includegraphics[width=5in]{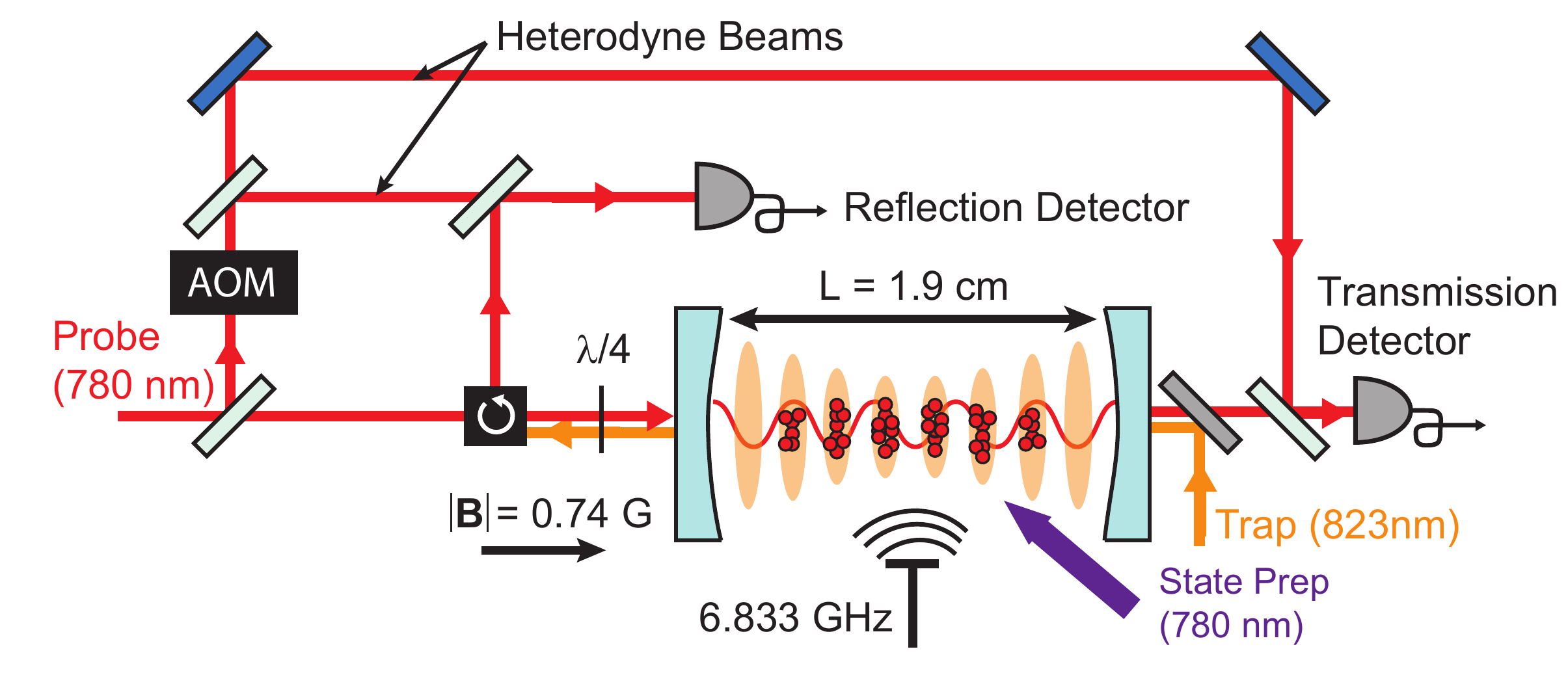}
\caption{{\bf Simplified experimental diagram.} The one dimensional optical lattice trap at $\lambda_l = 823$ nm is formed from a standing wave in the cavity (orange). We load $N_\circ = 4.0\times10^4$ to $7.2\times10^5$ $^{87}$Rb atoms into the trap and cool them to 10 $\mu$K.  The atomic sample extends $\sim 1$ mm along the axis of the $L = 1.9$ cm long optical cavity. State preparation is performed using a combination of 780 nm light (purple) for optical pumping and coherent ground state rotations performed with $6.833$ GHz microwaves from the dipole antenna. A uniform magnetic field is applied to provide a quantization axis and spectrally resolve the ground state zeeman sub-levels. The probe electric field forms a standing wave in the cavity, represented by the sinusoidal red line. The atom-cavity system is probed with 780 nm light (red), set to $\sigma^+$ polarization before entering the cavity. The probe light is separated from trap light using a dichroic mirror (grey) on the probe transmission side. The probe light is detected in both reflection and transmission with a heterodyne beam frequency shifted by an AOM.}
\end{figure}

\begin{figure}
\centering
\includegraphics[width=3in]{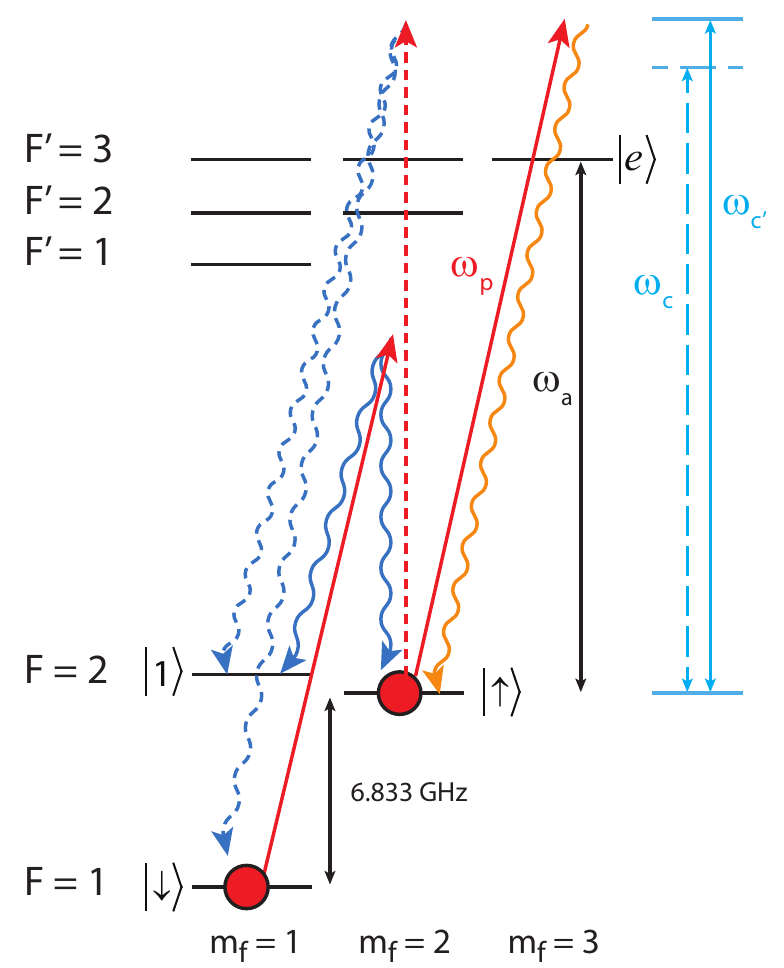}
\caption{{\bf Atomic energy level structure.} The relevant energy level structure of the $5\,^2$S$_{1/2}$ to $5\,^2$P$_{3/2}$ transition in $^{87}$Rb. The cycling transition has an optical atomic resonance frequency $\omega_a$. The cavity resonance with no atoms present (dashed light blue) with frequency $\omega_c$ is detuned to the blue of atomic resonance. The atom-cavity coupling creates a dressed cavity resonance (light blue) with frequency $\omega_{c'}$ which we probe using $\sigma^+$ laser light at frequency $\omega_p$ (red).  The cycling nature of the transition means scattering primarily maintains population in $\ket{\uparrow}$ (orange). Scattering of the probe light from atoms in $\ket{\downarrow}$, detuned by $\sim 6.8$ GHz, provides the fundamental limit to the cycling transition, as atoms can scatter to both $\ket{\uparrow}$ and $\ket{1} \equiv \ket{F=2,m_f=1}$ (dark blue).  Furthermore, imperfect polarization can lead to transitions for atoms in $\ket{\uparrow}$ to other internal states. An example of scattering from a $\pi$-polarized component of the probe is shown as the dashed red lines, with the state changing transitions in dark blue.
}
\end{figure}

\begin{figure} 
\centering
\includegraphics[width=5in]{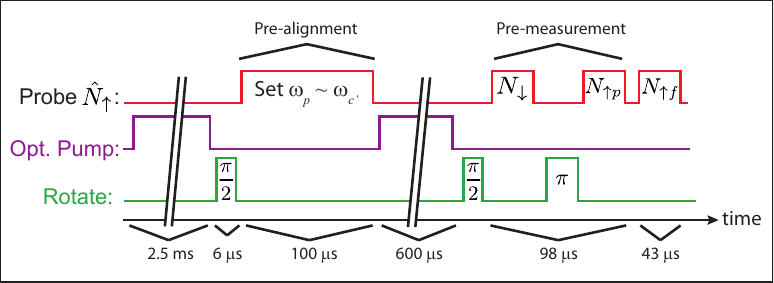}
\caption{{\bf Timing Sequence. } The experimental timing sequence showing probe laser pre-alignment and the pre-measurement that prepares a conditionally spin-squeezed state, followed by a final measurement to quantify the reduction in spin noise. Each optical pumping step (purple) prepares the ensemble in $\ket{\downarrow}$. Rotations (green) are performed by coupling $\ket{\uparrow}$ and $\ket{\downarrow}$ with a coherent microwave source. Probe laser pulses (red) correspond to individual measurements $\hat{N}_\uparrow$, with the measurement outcomes labeled as shown.} 
\end{figure}


\begin{figure}
\centering
\includegraphics[width=3.5in]{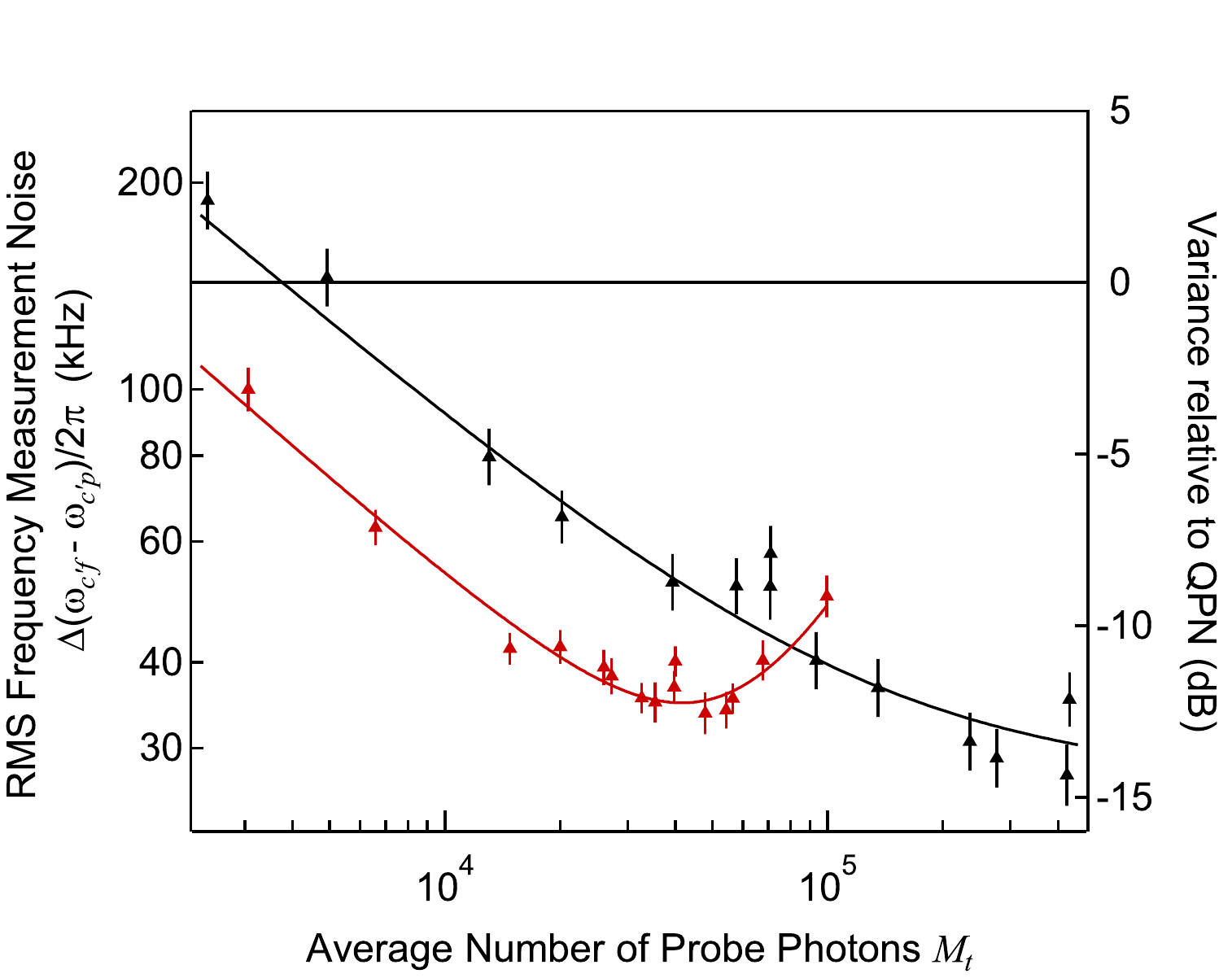}
\caption{{\bf Comparing measurement noise with and without atoms.} The fluctuations in the difference between two measurements of the dressed cavity frequency $\omega_{c'f}-\omega_{c'p}$, used to determine the difference of population measurements $N_{\uparrow f} - N_{\uparrow p}$, is plotted versus the average number of probe photons $M_t$. On the left axis, the fluctuations are expressed as the standard deviation $\Delta (\omega_{c'f}-\omega_{c'p})/2\pi$ in absolute frequency units. On the right axis, the same fluctuations are expressed as the ratio of the variance $(\Delta (\omega_{c'f}-\omega_{c'p}))^2$ to the variance $(\Delta \omega_{c'p})^2$ caused by the quantum projection noise (QPN) of a CSS. For the ensemble of $N = 4.8\times10^5$ atoms here, QPN causes fluctuations $\Delta \omega_{c'} = 2\pi\times 144(9)$ kHz, indicated by the line at 0 dB. Measurement noise is compared with (red) and without (black) the atoms loaded in the trap. For the case with atoms, the right axis is equivalent to $R$. The lines are fit to the data.  The error bars are 1 std. dev.
}
\end{figure}
\begin{figure}
\centering
\includegraphics[width=3.5in]{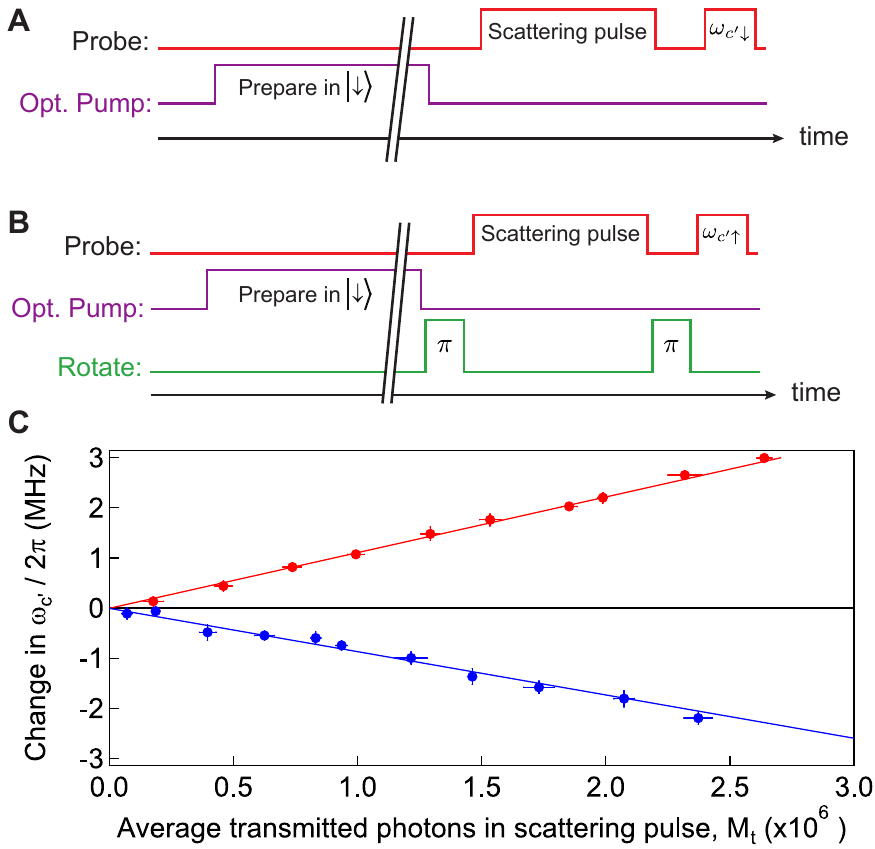}
\caption{{\bf Probe-induced population change. } ({\bf A}) To measure the sum of transition probabilities $p_{\downarrow \uparrow} + p_{\downarrow 1}$, we first prepare atoms in $\left|\downarrow\right\rangle$ with optical pumping (purple). Next, a scattering probe pulse (red), quantified by the average number of transmitted probe photons $M_t$, causes some atoms to change state to $\ket{\uparrow}$ and $\ket{1}$. Many photons per atom in $\ket{1}$ are scattered into free space, allowing the atoms in $\ket{1}$ to transition to $\ket{\uparrow}$. Thus, the measurement of the dressed cavity frequency $\omega_{c' \downarrow}$ gives the total number of atoms scattered out of $\left|\downarrow\right\rangle$.
({\bf B}) To measure the transition probability $p_{\uparrow \downarrow}$, we prepare atoms in $\left|\uparrow\right\rangle$ using optical pumping and a microwave $\pi$ pulse (green). The imperfection in the $\sigma^+$ polarized probe used for the scattering pulse causes some atoms to change state to $\ket{\downarrow}$. We again assume all atoms that scatter to $\ket{1}$ immediately transition back to $\ket{\uparrow}$. We swap the populations in $\ket{\uparrow}$ and $\ket{\downarrow}$ with another microwave $\pi$ pulse, so the measurement of $\omega_{c' \uparrow}$ gives the number of atoms that scattered to $\left|\downarrow\right\rangle$.
({\bf C}) Measurements of the dressed cavity frequency due to probe-induced internal state-changing transitions, with $\omega_{c' \downarrow}$ (red) described in {\bf A}, and $\omega_{c' \uparrow}$ (blue) described in {\bf B}. The lines are fits to the data, yielding a change in $\omega_{c'}$ per transmitted photon $\delta \omega_{c' \downarrow} = 2\pi \times 1.11(2)$  Hz/photon (red) and $\delta \omega_{c' \uparrow} = -2\pi \times 0.86(5)$ Hz/photon (blue). Here $N = 2.1\times10^5$ atoms.
}
\end{figure}

\clearpage


 \begin{table}
 \centering 
\begin{tabular}{ |p{6cm}||p{4cm}|}
 \hline
 Noise Source & $R^{-1}$ (uncertainty)    \\
 \hline
 Observed Optimum               & 16(2)       \\
 Photon Shot Noise $r_{PSN}$    & 32(4)		\\
 Technical Noise Floor $R_t$    & 73(34)	     \\
 \hspace{5mm} Laser Linewidth   & 520(250)     \\
 Classical Noise $r_c$	        & 67(15)             \\
 \hspace{5mm} Variable Damping $R_o$  & 620(60)               \\
 \hspace{5mm} Photon Recoil  $R_{ext, c}$    & $4.5(5)\times10^3 $         \\
 \hspace{5mm} Population Change $R_{pop,c}$  & $3.2(4)\times10^4$ \\
 Quantum Noise $r_q$            &$> 7\times10^5$     \\
 \hspace{5mm} Photon Recoil $R_{ext, q}$    & $4.8(5))\times10^5$     \\
 \hspace{5mm} Population Diffusion $R_{pop,q}$ & 1700(300)  \\
  \hline
 \end{tabular}
  \caption{Contributions to the observed spin noise reduction $R$ from a fit to the data versus probe strength $M_t$ that includes photon shot noise, a noise floor, quantum noise, and classical noise. Each value is given at the optimum spin noise reduction at $M_t = 4.1\times10^4$.  The indented rows show the calculated contributions from the various noise sources to each term. The uncertainties are 68.3\% confidence intervals.} 
 \end{table}


\end{document}